# Self-protection and self-healing in the context of cognitive radio

Mohammed Zakarya Baba-Ahmed, Badr Benmammar
Dept. Electrical and Electronic Engineering
Laboratory of Telecommunication of Tlemcen, LTT
Tlemcen, Algeria
{zaki.babaahmed, badr.benmammar}@gmail.com

Fethi Tarik Bendimerad
Dept. Electrical and Electronic Engineering
Laboratory of Telecommunication of Tlemcen, LTT
Tlemcen, Algeria
ftbendimerad@gmail.com

*Abstract*—Cognitive Radio (CR) operates in different fields as varied, one of these is cognitive radio networks. In this paper, we propose a new approach used CR, which aims to manage potential failures of computer systems and applications through the introduction of two aspects of autonomous networks to make systems capable of managing themselves with minimum human intervention.

## I. INTRODUCTION

The idea of cognitive radio was officially presented by Joseph Mitola III in a seminar at KTH, the Royal Institute of Technology in 1998, later published in an article by Mitola and Gerald Q. Maguire in 1999 [1].

The term cognitive radio was frequently used to refer to a system able to recognize its environment and to take advantage of this information. Sometimes, it is considered more restrictive as a system with high frequency agility to explore opportunities that may exist in the frequency spectrum [2].

CR provides a good spectrum management occupy or operate in the unoccupied bands of radio spectrum, and of course, thus improving spectrum management. And that this is due to the Software Defined Radio (SDR).

The SDR is a radio communication system which can adapt to any frequency band and receive any modulation using the same material [1].

Autonomous networks are mainly represented by the self-management, which aims to become less dependent computer systems users. The latter is characterized by four distinct points: Self-optimization, self-configuration, self-protection and self-healing.

An autonomous system will never settle for the present situation. It will be constantly monitoring the achievement of predefined system or performance levels to ensure that all systems are running at an optimal level, it must be able to install and configure the software automatically. It must also identify, detect, and protect valuable corporate resources from many threats, as he will be able to find and repair any problems to ensure that the systems function properly [3].

In this paper we begin by establishing the link between CR and autonomous network by integrating the specifications of self-management in cognitive systems, then we present an approach to manage potential interference problems may appear between a Primary User (PU) with a license on the spectrum and a Secondary User (SU) will allocated channels in the spectrum.

## II. AUTONOMY IN THE CONTEXT OF CR

Autonomy in CR networks is mainly focus on spectrum management which allows the improvement of the throughput without degrade the communications of others. Several studies were presented on the different characteristics of self-management:

### A. Self-optimization of cognitive engine

This work has been already achieved by [4] which allow complete and autonomous adaptation of cognitive engine to:

- Respect the regulatory framework that controls access to the spectrum ;
- Meet the needs of the user in terms of quality of service ;
- Let's make an optimized management of available resources [4].

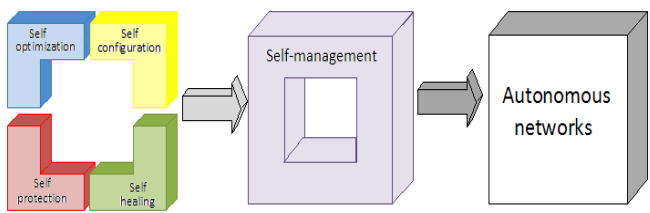

Figure 1. The four basic elements

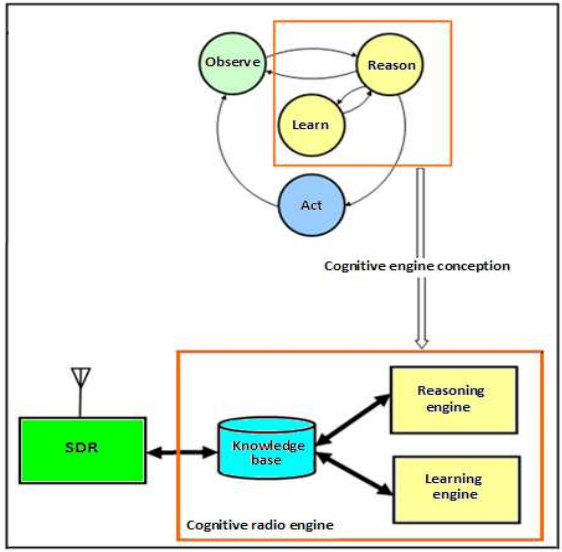

Figure 2. Self-optimization of cognitive engine

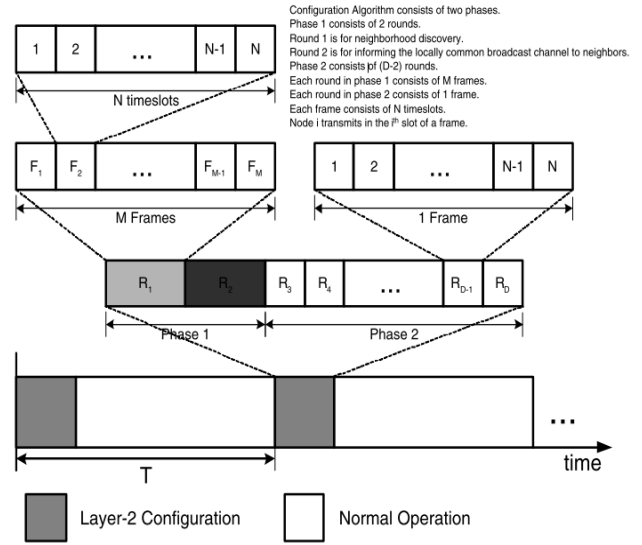

Figure 3. Operation cycle of a diameter-aware node.

In Figure 2, the elements of CR communicate with SDR support through the cognitive engine. This latter is responsible of the optimization and control of the SDR module based on some input part parameters such as information from sensory perception to reasoning with the system or the learning of the radio environment, user context and network status.

Here, the knowledge base keeps the system states and actions available. The reasoning engine uses the knowledge base to select the best action. The Learning Engine performs the manipulation of knowledge based on the observed information (information on channel availability, the error rate in the channel, etc.) [1].

### B. Layer-2 Self-Configuration Algorithms

In a network with compatible devices, self-configuration Layer 2 in the CR involves determining a common set of channels to facilitate communication between participating nodes. It is a unique challenge because maybe the CR network nodes ignore:

- Their neighbors ;
- The channels on which they can communicate with a neighbor.

The authors in [5] have proposed a CR network algorithm with a distributed time-efficient for self-configuration of Layer 2.

During the layer-2 auto-configuration process, the Time Division Multiple Access (TDMA) scheme is used for communication among nodes. Time is divided into O(D) rounds. A round is defined as the time taken for every node to communicate with each one of its neighbors using a (local) broadcast mechanism. Each round consists of equal-sized intervals referred to as frames. The number of frames in a round may vary as shown in Figure. 3. A round in phase 1 consists of M frames while a round in phase 2 consists of only one frame. Each frame is further divided into N timeslots, each of equal length. Node $i$ transmit during the $i$th timeslot in each frame (see Figure. 3) and all other nodes are in receiving mode during the $i$th timeslot. We next present more details of the algorithm [5].

### C. Self-Awareness in cycle of cognition

Self-awareness demonstrates the development of concepts that constitute a major problem in the design of cognitive self-managed networks according to the vision of the Future Internet networks being developed in the Self-NET project [6].

Figure 4 unifies and makes some of the key issues for the development of cognitive systems in self-awareness and currently performs the implementation of these with real test beds that will reveal the practical implementation and coordination problems in the deployment of cognitive cycles for use cases and application of the theoretical framework presented in Figure 4.

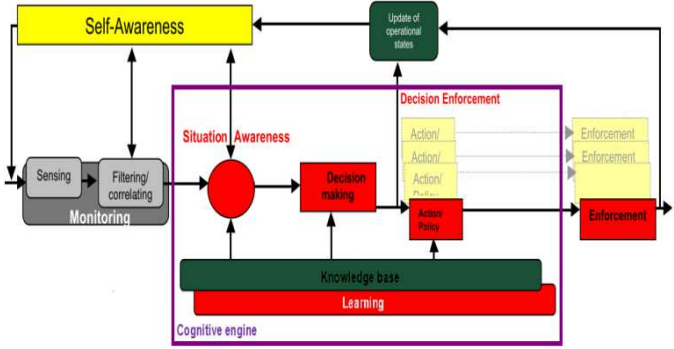

Figure 4. Self-NET logical architecture of the cognitive cycle for self-management [6]

## III. PROPOSES SOLUTION

As you've probably noticed, we have not discussed two characteristics of autonomous systems, self-protection and self-healing. Our approach in this paper will focus on these two points. But for now we'll see what kind of transmission will be dedicated this autonomy.

Quality of Service (QoS) is part of the complexity of data transmission, as Bandwidth, Delay, Loss and Jitter. What better than to ensure the transmission of the video conference who has a very high priority to insure in CR.

TABLE I. QOS NETWORKS DATA TYPE AND SENSITIVITIES

| Transmission Type | characteristic | | | |
|---|---|---|---|---|
| | *Bandwidth* | *Delay* | *Loss* | *Jitter* |
| Voice | 1 | 4 | 3 | 4 |
| E-Commerce | 2 | 4 | 4 | 2 |
| Transactions | 2 | 4 | 4 | 2 |
| E-mail | 2 | 2 | 4 | 2 |
| Telnet | 2 | 3 | 4 | 2 |
| Casual browsing | 2 | 3 | 3 | 2 |
| Serious browsing | 3 | 4 | 4 | 2 |
| File transfers | 4 | 2 | 3 | 2 |
| Video conferencing | 4 | 4 | 3 | 4 |
| Multicasting | 4 | 4 | 4 | 4 |

Table I illustrates the varied application data types and their respective network sensitivities regarding bandwidth, packet loss, delay. The characteristics of the data types are denoted as very high – 5, high – 4, medium – 3, low – 2, and very low – 1 [7].

### A. Proposes architecture

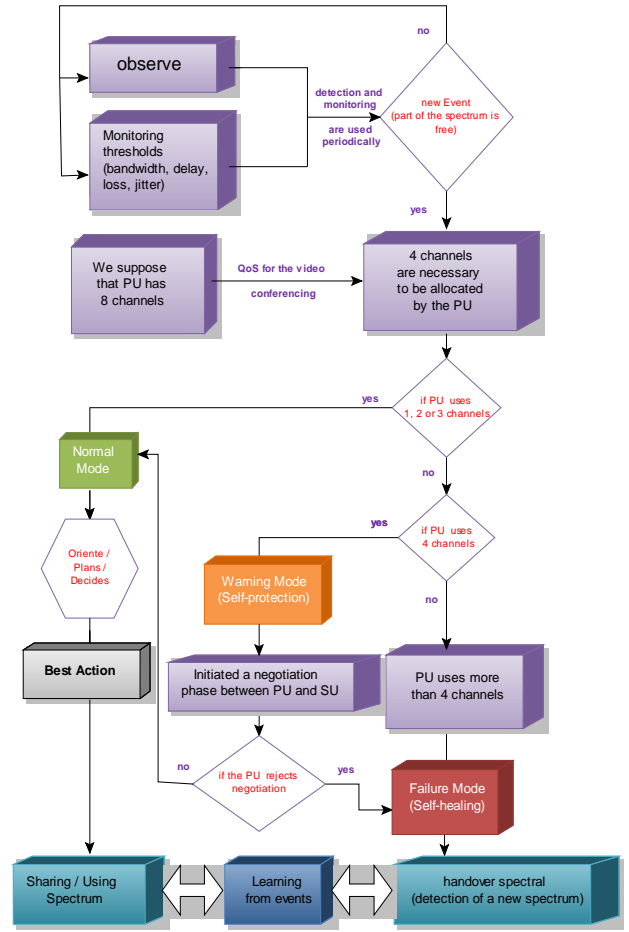

Figure 5. Self-protection and self-healing of a cognitive radio node to management failures

### B. Principle

The principle of this approach is that a SU will detect and monitor the QoS thresholds periodically to find a free part of a spectrum licensed of a PU. It is assumed in this case that the PU has 8 channels and thus to ensure the QoS of the video conference with a sensitivity rated 4 (i.e. high); the SU will allocate 4 of these 8 channels with the agreement of PU.

Now the PU has three possible scenarios:

1/ If the PU uses one, two or three of its channels, the SU is in *Normal Mode*. So the SU directs, plans and decides the best action to take after the SU uses/share the spectrum with the PU.

2/ If the PU uses 4 channels available, in this case we are in *Warning Mode*, using its 8 channels can cause interference between the SU and the PU during transmission, which will seriously disturb the SU as it is supposed guaranteed QoS for video conferencing, then a negotiation phase will be handled between the PU and SU is the tamper, after negotiation we would have two possible cases:

- The first case will occur if the PU cooperates with the SU and agrees to assign at least one

transmission channel, so the SU will return to the first scenario (i.e. in *Normal mode*).

- The second case occurs if the PU refuses any type of negotiation, the SU will entered a new phase that is mentioned in the third scenario.

3/ If the PU decides to use more than four channels, in this case, not enough channels to allocated by the SU, it is in the *Failure Mode*, the interference will be guaranteed and no negotiation phase will be given, so a new phase is required, this is called the Self-healing. In this phase we made a changing spectrum or in other words we need a *Spectral handover*.

*Auto-protection* is generated by the negotiation phase that seeks a dynamic spectrum access; there are various existing approaches for dynamic spectrum allocation:

- Medium Access Control (MAC);
- Local Negotiation;
- Game Theory;
- Multi Agent Systems (MAS);
- MARKOV's chains.

The latter approach will be more conducive to the negotiation of our work because normally the most known approaches based on MARKOV's chains; the authors model the interactions between users and calculate the probabilities of blocking and non-completion as the main parameters evaluation [8].

The *Self-healing* is caused by the phase of the changing spectrum (*Spectral Handover*) to conduct a new detection of another spectrum with the same requirements in terms of quality of service for video conferencing.

Learning is done by the events that occurred during the three modes and the results obtained by the management of the two phases, which leads to enrich the knowledge base by *Automatic Learning*.

## IV. CONCLUSION

In this paper we presented several parts of the self-management link cognitive radio with the autonomous networks, but our work was mainly focused on two parts of autonomy, of course we speak about self-protection and self-healing of a cognitive radio node on which we studying the case of maturity or non-cooperation between primary and secondary user with the use of negotiation in case of threat of interference between the latter two, and in case of changing spectrum caused by out of cooperation.

In our future work, we will seek to minimize the failure rate negotiation between PUs and SUs whom coexist in the same spectrum and we will study the impact of self-management in cognitive radio networks.